\documentclass[copyright,creativecommons]{eptcs}
\providecommand{\event}{ICE 2016} 
\newcommand{\mse}{\texttt{monitored-session-erlang}}
\usepackage[scaled=.7]{beramono}
\usepackage[T1]{fontenc}
\usepackage{color}
\usepackage{amsmath,amssymb,centernot,colortbl,mathwidth,scalerel,stmaryrd,url,xspace,mathrsfs,bm}
\usepackage{tensor}
\usepackage[table]{xcolor}
\usepackage{longtable}
\usepackage{mathtools}
\usepackage{amsthm}

\theoremstyle{definition}

\usepackage{hyperref}
\hypersetup{colorlinks=true,citecolor=blue}

\usepackage[numbers,sort&compress,sectionbib,square]{natbib}

\usepackage{mathpartir}
\mprset{sep=1em}

\usepackage{float}
\floatstyle{boxed}
\restylefloat{figure}
\usepackage{verbatim}

\newcommand\doubleplus{+\kern-1.3ex+\kern0.8ex}

\definecolor{shade}{RGB}{223,223,223}

\newcommand{\ba}{\begin{array}}
\newcommand{\ea}{\end{array}}

\newcommand{\bl}{\ba[t]{@{}l@{}}}
\newcommand{\el}{\ea}

{\[\begin{array}{@{}lr@{\hspace{2px}}c@{\hspace{2px}}l}\ignorespaces}
{\end{array}\]\ignorespacesafterend}

\usepackage[T1]{fontenc}

\usepackage{listings}
\DeclareMathAlphabet{\mathpzc}{OT1}{pzc}{m}{it}

\usepackage{caption}

\usepackage{subcaption}

\usepackage{ragged2e, array}
\usepackage{microtype}
\usepackage{setspace}

\makeatletter
\renewcommand{\paragraph}{%
  \@startsection{paragraph}{4}%
  {\z@}{0ex \@plus 1ex \@minus .2ex}{-1em}%
  {\normalfont\normalsize\bfseries}%
}
\makeatother

\usepackage{multicol}
\usepackage{vwcol}

\usepackage{paralist}
\usepackage{csquotes}
\usepackage{breakurl}
\usepackage{wrapfig}
\usepackage{listings}
\usepackage{upquote}
\usepackage{color}

\definecolor{bluekeywords}{rgb}{0.13,0.13,1}
\definecolor{greencomments}{rgb}{0,0.5,0}
\definecolor{turqusnumbers}{rgb}{0.17,0.57,0.69}
\definecolor{redstrings}{rgb}{0.5,0,0}

\lstdefinelanguage{Scribble}{
  morekeywords={
  	global, protocol, role, from, to, interruptible, with, do, instantiates, par, and, rec, continue, choice, at, initiates, handle, returning, call, local, or, new
  },
  otherkeywords={ },
  keywordstyle=\color{bluekeywords},
  sensitive=true,
  basicstyle=\linespread{0.9}\ttfamily,
	breaklines=true,
  xleftmargin=\parindent,
  belowskip=\bigskipamount,
  aboveskip=\bigskipamount,
  tabsize=4,
  morecomment=[l][\color{greencomments}]{///},
  morecomment=[l][\color{greencomments}]{//},
  morecomment=[s][\color{greencomments}]{{(*}{*)}},
  morestring=[b]",
  showstringspaces=false,
  literate={`}{\`}1,
  frame=none,
  showlines=false,
  frame=single,
  stringstyle=\color{redstrings},
}

\providecommand*{\doi}[1]{\href{http://dx.doi.org/#1}{doi: #1}}

\title{An Erlang Implementation of Multiparty Session Actors}
\author{Simon Fowler
\institute{The University of Edinburgh\\ Edinburgh, UK}
\email{simon.fowler@ed.ac.uk}
}
\def\titlerunning{An Erlang Implementation of Multiparty Session Actors}
\def\authorrunning{S. Fowler}
\begin{document}
\maketitle

\begin{abstract}
By requiring co-ordination to take place using explicit message passing instead of relying on shared memory, actor-based programming languages have been shown to be effective tools for building reliable and fault-tolerant distributed systems. Although naturally communication-centric, communication patterns in actor-based applications remain informally specified, meaning that errors in communication are detected late, if at all.

Multiparty session types are a formalism to describe, at a global level, the interactions between multiple communicating entities.
This article describes the implementation of a prototype framework for monitoring Erlang/OTP \texttt{gen\_server} applications against multiparty session types, showing how previous work on multiparty session actors can be adapted to a purely actor-based language, and how monitor violations and termination of session participants can be reported in line with the Erlang mantra of `let it fail'. Finally, the framework is used to implement two case studies: an adaptation of a freely-available DNS server, and a chat server.
\end{abstract}

\section{Introduction}
Programming concurrent and distributed systems is a challenge---the introduction of concurrency and distribution introduces issues such as deadlocks, race conditions, and node failures; issues which simply do not arise when developing single-threaded applications.

The actor model is a model of concurrency introduced by~\citet{hewitt:actors} and discussed in the context of distributed computing by~\citet{agha:actors}. \emph{Actors}---entities with a unique ID and a message queue called a mailbox---react to incoming messages in three ways: by asynchronously sending a finite set of messages to other actors; spawning a finite number of new actors; and changing how they react to future messages.

The actor model provides an ideal theoretical basis for programming languages designed for programming robust and fault-tolerant distributed systems. Programming languages such as Erlang~\cite{armstrong:erlang} and Elixir take actors as primitive entities, implemented as lightweight processes which communicate only through explicit message passing. By eschewing shared memory as a method of co-ordination, applications are naturally built in a style fostering fault isolation, scalability, and modularity. Of particular note is Erlang's approach to failure handling---an approach succinctly described as `let it fail', wherein processes should terminate upon encountering a fault, and be restarted by a supervisor process. 

Moving to a communication-centric programming paradigm has its own issues, however: in particular, how do we document the communication patterns expected within an application, and how do we ensure that the application conforms to these communication patterns? \emph{Multiparty Session Types}~\cite{honda:multiparty} are a type formalism in which a \emph{global type} describing interactions between participants in a session---a series of interactions between participants---can be projected into \emph{local types} describing the interaction from the point of view of an individual participant. It is then possible to check conformance either statically using a type checker, or dynamically through runtime monitoring techniques.

We build upon the work of~\citet{neykova:actors}, who describe a conceptual framework in which actors are treated as entities which are simultaneously involved in multiple possibly-interacting sessions. Their conceptual framework is realised as a library building upon the Cell actor framework in Python. 

While actors can be emulated, Python remains an imperative language with mutability and shared memory concurrency. 
This article seeks to demonstrate the applicability of this conceptual framework in Erlang, which as well as taking actors as primitive, is a functional language which forbids co-ordination via shared memory. In addition, we investigate how dynamic monitoring of communication against multiparty session types may be integrated with the `let it fail' methodology of Erlang.

\subsection{Contributions}
The contributions of this article are:

\begin{compactitem}
\item A tool, \mse{}\footnote{\mse{} can be found online at \url{http://www.github.com/SimonJF/monitored-session-erlang}. Chat server: \url{http://www.github.com/SimonJF/mse-chat}, DNS server: \url{http://www.github.com/SimonJF/erlang-dns}.},
for monitoring communication in Erlang applications, based on the session actor framework of~\citet{neykova:actors}. The implementation uses native actor functionality as opposed to emulating actors using AMQP and Python Greenlets, and has a simpler invitation workflow as a result. Additionally, we allow actors to take part in multiple \emph{instances} of a session, which is necessary to implement server applications.
\item An extension of the Scribble protocol description language to use \emph{subsessions}~\cite{demangeon:subsessions} for introducing participants midway through a session, and for structuring possibly-failing sessions.
\item A discussion of ways to detect and act upon failures, including a two-phase commit to ensure messages are accepted in a multicast, and reachability analysis to ascertain whether a failed participant is required in the remainder of a session.
\item Two larger examples of session-based communication using the framework: a chat server, and an adaptation of a freely-available DNS server.
\end{compactitem}
\section{An Overview of Multiparty Session Actors}

Multiparty session types are a formalism to allow a protocol---a series of typed interactions between multiple participants---to be described as a \emph{global type}. As the name suggests, global types are a \emph{global} formalism, describing all of the interactions in that particular protocol. Global types can be projected into \emph{local} types, which describe the protocol from the point of view of a single participant.

In the traditional view of runtime monitoring of multiparty session types, each process is monitored by a single monitor, which checks incoming and outgoing messages to see whether they conform to a local type.
The actor model, on the other hand, lends itself to a different model of monitoring. Actors are naturally event-driven: upon processing a message from a mailbox, an actor can send a finite set of messages; spawn a finite set of new actors; and change the way it behaves upon encountering the next message. 
A more appropriate style of monitoring pioneered by~\citet{neykova:actors} is to treat actors as entities that can take part in multiple sessions. The core ideas behind multiparty session actors are that:

\begin{compactitem}
\item Actors may be involved in multiple sessions simultaneously.
\item Actors may play multiple roles in each session.
\item A message received in the course of one session may trigger an interaction in another session.
\end{compactitem}

In this setting, actors become containers for monitors and handlers for incoming session messages. Actors may be involved in multiple roles in multiple sessions, with the ability to co-ordinate between them, and with actor-wide state shared amongst the handlers for each session.

To describe protocols, we use Scribble~\cite{yoshida:scribble}, a human-readable protocol description language based on the theory of multiparty session types. Scribble is realised as a Java-based toolchain\footnote{\url{http://www.github.com/scribble/scribble-java}}, including components for parsing, validating well-formedness, and creating local projections of global types. 

\subsection{A Chat Server}\label{sec:overview:chat-server}
We illustrate the main concepts of the multiparty session actor method of designing actor-based applications in the context of our Erlang implementation, \mse{}.

\begin{wrapfigure}[10]{l}{8.5cm}
\includegraphics[width=\linewidth]{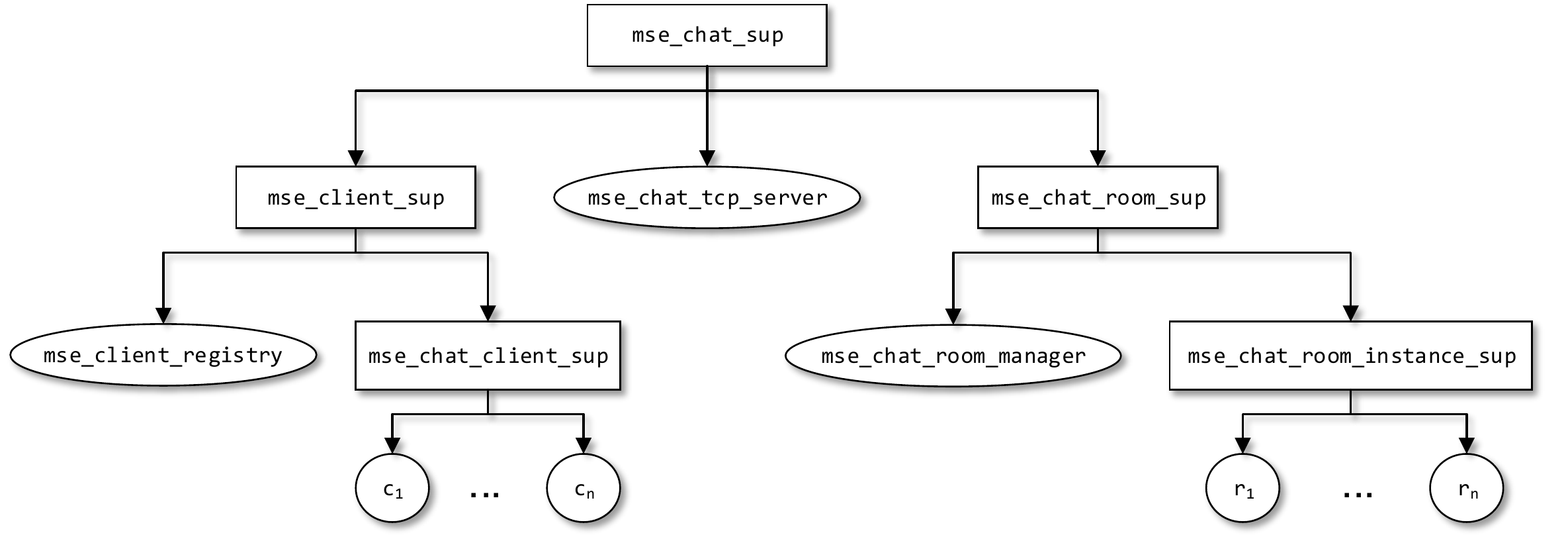}
\caption{Supervision Tree for Chat Server}
\label{fig:mse-chat-tree}
\end{wrapfigure}

Erlang/OTP applications are structured for reliability using a design pattern known as \emph{supervision hierarchies}, where workers (actors which perform computations) can be restarted by \emph{supervisors} if they terminate. 
Instead of attempting to recover from an unexpected or erroneous state, actors are designed to terminate and be restarted. 
Figure~\ref{fig:mse-chat-tree} shows the supervision tree for the chat system. The nodes in boxes denote supervisor actors, which do not participate in interactions themselves, but restart their child actors should the child actors terminate. Of interest to the protocol are three actors: \texttt{mse\_chat\_client} ($c_1, \ldots, c_n$ in the diagram), representing a chat client actor; \texttt{mse\_chat\_room\_instance} ($r_1, \ldots, r_n$ in the diagram), representing an instance of a chat room; and \texttt{mse\_chat\_room\_manager}, which maintains a registry of chat rooms.

The \texttt{mse\_chat\_tcp\_server} actor listens on a socket and accepts new clients, spawning a new \texttt{mse\_chat\_\ client} to handle requests from the new client. A client can either create or join a room by sending a message to \texttt{mse\_chat\_room\_manager}, which checks whether the room exists. Once a client is registered with the room, any chat messages sent should be distributed to all other participants in the room. The client can leave the session at any time and should be deregistered from any rooms to which it is registered.

\begin{figure}[htpb]
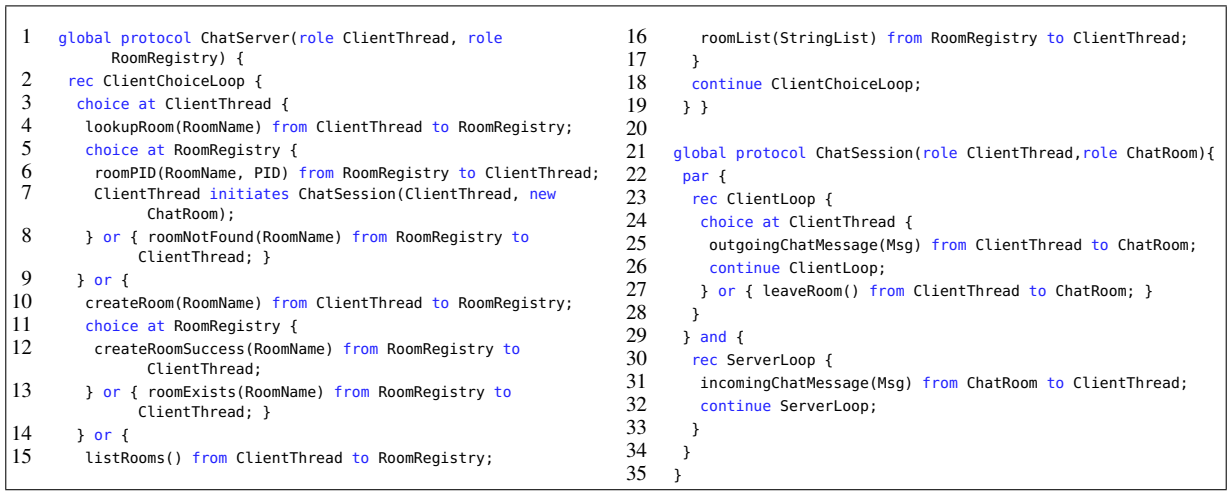

\begin{lstlisting}[numbers=left,xleftmargin=3em, multicols=2,frame=none]
global protocol ChatServer(role ClientThread, role RoomRegistry) {
 rec ClientChoiceLoop {
  choice at ClientThread {
   lookupRoom(RoomName) from ClientThread to RoomRegistry;
   choice at RoomRegistry {
    roomPID(RoomName, PID) from RoomRegistry to ClientThread;
    ClientThread initiates ChatSession(ClientThread, new ChatRoom);
   } or { roomNotFound(RoomName) from RoomRegistry to ClientThread; }
  } or {
   createRoom(RoomName) from ClientThread to RoomRegistry;
   choice at RoomRegistry {
    createRoomSuccess(RoomName) from RoomRegistry to ClientThread;
   } or { roomExists(RoomName) from RoomRegistry to ClientThread; }
  } or {
   listRooms() from ClientThread to RoomRegistry;
   roomList(StringList) from RoomRegistry to ClientThread;
  }
  continue ClientChoiceLoop;
 } }

global protocol ChatSession(role ClientThread,role ChatRoom){
 par {
  rec ClientLoop {
   choice at ClientThread {
    outgoingChatMessage(Msg) from ClientThread to ChatRoom;
    continue ClientLoop;
   } or { leaveRoom() from ClientThread to ChatRoom; } 
  }
 } and {
  rec ServerLoop {
   incomingChatMessage(Msg) from ChatRoom to ClientThread;
   continue ServerLoop;
  }
 }
}
\end{lstlisting}
\caption{Global Protocols for Chat Server}
\label{fig:scribble-chatsession}
\end{figure}

Figure~\ref{fig:scribble-chatsession} shows two Scribble protocols describing the chat server.
The \texttt{ChatServer} protocol describes the interactions between a client and the room registry before joining a room. Interactions are of the form \lstinline+messageName(Payload) from FromRole to ToRole1, ..., ToRoleN+, meaning that the role with name \texttt{FromRole} sends a message with name \texttt{messageName} and payload type \texttt{Payload} to roles with names \texttt{ToRole1} \ldots \texttt{ToRoleN}.

A common pattern in Erlang involves sending process IDs. We make use of a small extension to Scribble to implement the theory of \emph{nested protocols}, or \emph{subsessions}~\cite{demangeon:subsessions} which allow a child session to be initiated with some participants invited from the current session, and some invited externally. This is implemented using the \texttt{initiates} construct (Line 7), stating that \texttt{ClientThread} starts the \texttt{ChatSession} protocol, externally inviting another actor to fulfil the \texttt{ChatRoom} role.

\begin{wrapfigure}[5]{l}{8.5cm}
\vspace{-1em}
\begin{lstlisting}[language=erlang,frame=none]
config() ->
  [{mse_chat_client, [{"ChatServer", ["ClientThread"]},
                      {"ChatSession", ["ClientThread"]}]},
   {mse_chat_room_manager, [{"ChatServer", ["RoomRegistry"]}]},
   {mse_chat_room_instance, [{"ChatSession", ["ChatRoom"]}]}].
\end{lstlisting}
\vspace{-1em}
\end{wrapfigure}
We begin by creating a configuration file associating roles with actors:
\texttt{mse\_chat\_client} can play \texttt{ClientThread} in both \texttt{ChatServer} and \texttt{ChatSession},  \texttt{mse\_chat\_room\_manager} plays \texttt{RoomRegistry} in \texttt{ChatServer}, and \texttt{mse\_chat\_room\_\ instance} plays \texttt{ChatRoom} in \texttt{ChatSession}.

Instead of using send and receive primitives directly, Erlang developers make heavy use of \emph{OTP behaviours}, which provide boilerplate functionality and require a developer to implement a number of callbacks in order to provide application logic. 
As an example, the \texttt{gen\_server} behaviour abstracts over Erlang's communication primitives to provide an event loop, invoking callbacks such as \texttt{handle\_cast} to process incoming messages. To participate in sessions in \mse{}, actors must implement the \texttt{ssa\_gen\_server} behaviour, described further in Section~\ref{sec:design-impl:ssa-gen-server}.

We begin by looking at the implementation of \texttt{mse\_chat\_room\_manager}. When an actor is spawned, the \texttt{ssactor\_init} callback is invoked with the arguments and an \emph{initiation key} for starting sessions, and should return the initial actor state. 
When an actor is invited to join a session, the \texttt{ssactor\_join} callback is invoked, and the return value defines whether the invitation is accepted or declined. The \texttt{ssactor\_handle\_message} callback is invoked when a session message has been received and accepted by the monitor.

\begin{figure}[htpb!]
\begin{lstlisting}[language=erlang,numbers=left,xleftmargin=3em, multicols=2,frame=none]
ssactor_init(_Args, _InitKey) ->
  #room_manager_state{rooms=orddict:new()}.
ssactor_join(_, _, _, State) ->
  {accept, State}.
ssactor_conversation_established("ChatServer", 
    "RoomRegistry", _CID, ConvKey, State) ->
  {ok, State}.
ssactor_handle_message("ChatServer", "RoomRegistry", 
    _, _, "lookupRoom", [RoomName], State, ConvKey) ->
  handle_get_room(ConvKey, RoomName, State),
  {ok, State}.

handle_get_room(ConvKey, RoomName, State) ->
  RoomDict = State#room_manager_state.rooms,
  case orddict:find(RoomName, RoomDict) of
    {ok, RoomPID} ->
     mse_chat_client:found_room_pid(ConvKey, 
        RoomName, RoomPID);
    error -> 
     mse_chat_client:room_not_found(ConvKey, RoomName)
  end.
\end{lstlisting}

\caption{Excerpt from \texttt{mse\_chat\_room\_manager} implementation}
\label{fig:mse-chat-manager-impl}
\end{figure}
\vspace{-1em}
Figure~\ref{fig:mse-chat-manager-impl} shows an excerpt from the implementation of \texttt{mse\_chat\_room\_manager}. The actor creates a new map structure in the \texttt{ssactor\_init} callback to hold rooms that will be created later; always accepts invitations to join sessions; and does nothing when the session is established. The \texttt{ssactor\_handle\_message} callback details how the \texttt{lookupRoom} message in the \texttt{ChatServer} protocol is handled: the actor queries the map between room names and PIDs, sending the PID to the requesting actor if it is found, and sending a ``room not found'' message if not. In line with Erlang development practices, message passing is abstracted as an API call: as an example, the \texttt{mse\_chat\_client:room\_not\_found} function is defined as
\lstinline[language=erlang]+room_not_found(ConvKey, RoomName) -> conversation:send(ConvKey, ["ClientThread"], "roomNotFound", [RoomName]).+

The \texttt{conversation:send} function sends a session message \texttt{roomNotFound} with payload \texttt{RoomName} to \\\texttt{ClientThread}; \texttt{ConvKey} (provided by \texttt{ssactor\_handle\_message}) is used to identify the correct monitor.
The implementation of the client follows a similar pattern. We show an excerpt of the implementation of the client in Figure~\ref{fig:mse-chat-client-impl}; note that a \texttt{ChatServer} session is initiated in \texttt{ssactor\_init}.

\begin{figure}[htpb!]
\begin{lstlisting}[language=erlang, multicols=2, frame=none, numbers=left,xleftmargin=3em]
ssactor_init([ClientID, ClientSocket], InitKey) ->
  State = 
   #client_state{client_id=ClientID, 
     client_name=undefined, client_socket=ClientSocket, 
     init_key=InitKey},
  conversation:start_conversation(InitKey, 
    "ChatServer", "ClientThread"),
  inet:setopts(ClientSocket, [{active, true}]),
  State.
  
ssactor_conversation_established("ChatServer", 
    "ClientThread", _CID, ConvKey, State) ->
  conversation:register_conversation(main_thread, ConvKey),
  {ok, State};


handle_message(Message, State) ->
  InitKey = State#client_state.init_key,
  SplitMessage = string:tokens(Message, ":"),
  [Command|PacketRemainder] = SplitMessage,
  NewState =
    if Command == "JOIN" ->
       [RoomName|_Rest] = PacketRemainder,
       conversation:become(InitKey, main_thread, 
         "ClientThread", join_room, [RoomName]),
       State;
       [ ... ]
    end,
  {noreply, NewState}

ssactor_become("ChatSession", "ClientThread", chat, 
               [Message], ConvKey, State) ->
  handle_chat(ConvKey, Message, State),
  {ok, State};
\end{lstlisting}
\caption{Excerpt from the \texttt{mse\_chat\_client} implementation}
\label{fig:mse-chat-client-impl}
\end{figure}
\noindent
Recall that the \texttt{mse\_chat\_client} participates in both the \texttt{ChatServer} and \texttt{ChatSession} protocols. Upon receiving messages from a chat client program, the process must ensure that a message is sent in the correct session. For example, a `create room' packet from the client application must be handled by \texttt{ChatServer}, whereas a `send chat message' packet must be handled by \texttt{ChatSession}. 

To accomplish this, we make essential use of the ability to switch between roles.
In our implementation (shown in Figure~\ref{fig:mse-chat-client-impl}), we can \emph{register} a session to a key (Line 13), and can use this key to switch to the session (Line 24). As an example, the \texttt{ClientThread} session is registered with \texttt{main\_thread} key, and switches to this session in order to send a \texttt{lookupRoom} packet to the room registry.
The \texttt{mse\_chat\_room\_instance} actor uses a similar approach to broadcast messages to the chat room.

\section{Design and Implementation of \mse{}}
\subsection{System Overview}
\begin{figure}[t]
\centering

\begin{subfigure}[t]{0.42\linewidth}
\centering
\includegraphics[width=\linewidth]{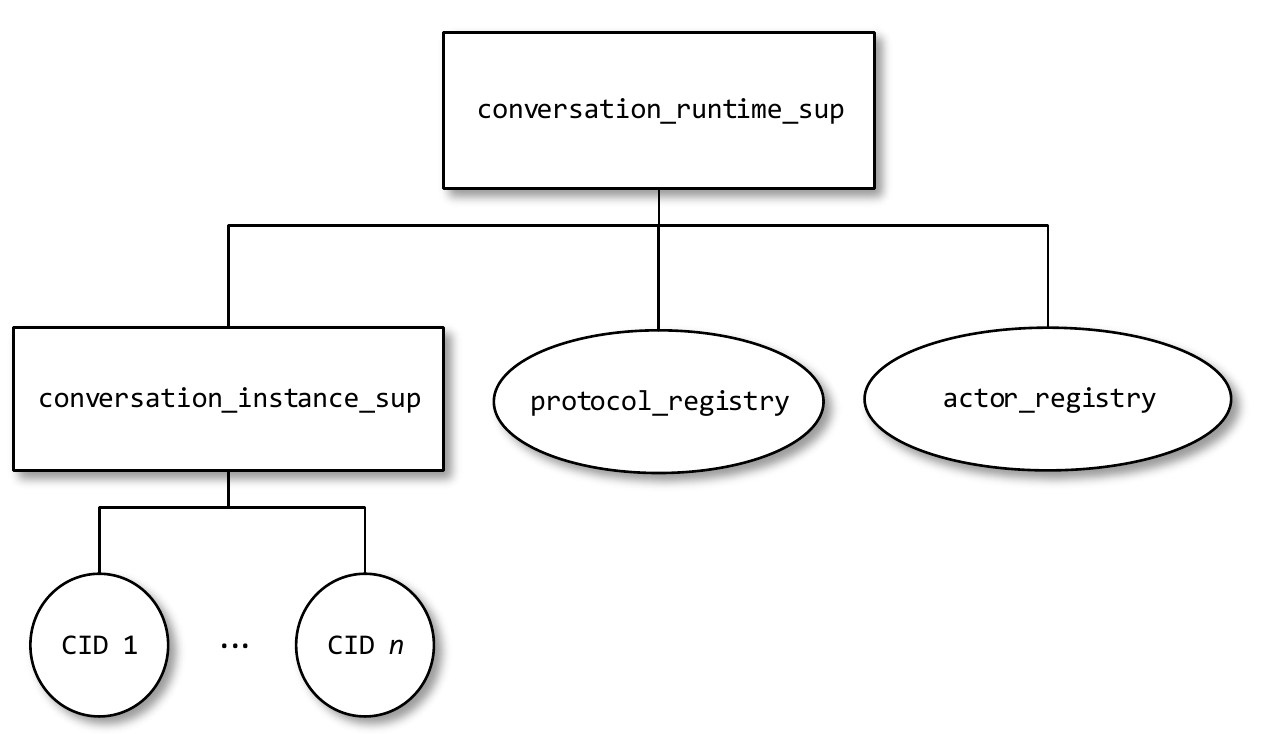}
\caption{\mse{} supervision tree}
\label{fig:mse-supervision-tree}
\end{subfigure}
~
\begin{subfigure}[t]{0.42\linewidth}
\centering
\includegraphics[width=\linewidth]{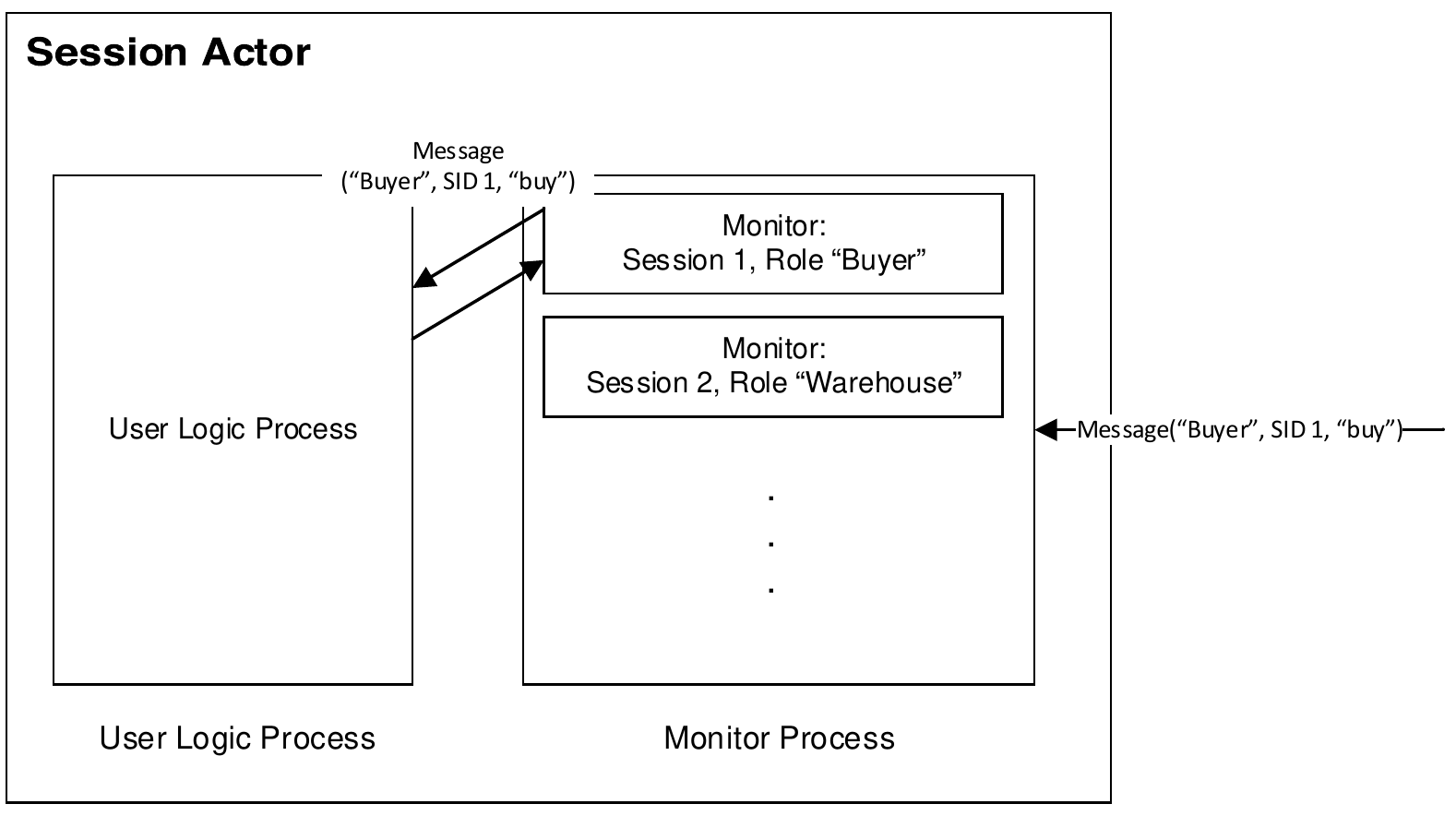}
\caption{Erlang Session Actors}
\label{fig:mse-session-actor}
\end{subfigure}

\caption{Components of \mse{}}
\label{fig:mse-components}
\end{figure}

The \mse{} system is implemented as an Erlang library. The supervision tree for the system is shown in Figure~\ref{fig:mse-supervision-tree}: \texttt{conversation\_runtime\_sup} is the root supervisor of the system, which restarts top-level processes should they fail. The \texttt{protocol\_registry} process associates protocols and roles with FSMs used for monitoring, and the \texttt{actor\_registry} process maintains a list of active actors which are registered to take part in sessions.

Each session is associated with a \emph{coordinating process} (depicted as \texttt{CID 1} \ldots \texttt{CID n} in Figure~\ref{fig:mse-supervision-tree}); coordinating processes are used to coordinate actions such as setting up a session and failure handling. Coordinating processes are arranged as children of the \texttt{conversation\_instance\_sup} process, but are not restarted should they fail. As a high-level overview, the system works as follows:

\begin{compactitem}
\item Local projections of protocols are used to generate monitors based on communicating finite-state machines (CFSMs).
A configuration file defines which roles, in which protocols, actors may fulfil.
\item When an actor is spawned, it is added to the \texttt{actor\_registry}, allowing it to participate in sessions.
\item A session \emph{initiator} begins a session. At this point, eligible actors are invited to fulfil each role in the protocol. When all roles are fulfilled, the participants of the session are notified that the session has been initiated successfully and are provided with the monitor FSM to use. 
Conversely, if it is not possible to fulfil all of the roles then all actors registered in the session are notified of the failure.
%
\item Session messages are sent using a session API, and processed by session actors. All communication using the session API is mediated by monitors, and messages which do not conform to the protocol are rejected, with an exception thrown in the sender.
\item Due to the supervision tree structure within the Erlang applications, we \emph{cannot assume that participants are alive for the duration of the session}. Consequently, we provide failure detection mechanisms, which detect when the session can no longer safely proceed.
%
\item When the session is over (or an actor ends it prematurely due to an error), a participant calls a function which notifies other participants in the session that the session has ended.
\end{compactitem}

\subsection{Erlang Session Actors}

A considerable insight due to Neykova and Yoshida is that instead of having a single monitor, session actors can fulfil multiple roles in multiple protocols.
At an abstract level, we can think of an Erlang session actor as containing seven components: a process ID, a term currently being evaluated, a mailbox, an actor state, a monitor lookup table, a routing table mapping session / role pairs to actor PIDs, and a message handler function. 
Each monitor can be uniquely identified using a pair of a session ID and a role. 

\paragraph{Role Registration}
\citet{neykova:actors} associate actors with roles using Python decorators, which is an appealing and intuitive method of associating message handlers, protocols, and roles together. It is unclear how it would be possible to allow multiple \emph{instances} of sessions, which is an important requirement when writing server applications. For example, in the chat server, a single \texttt{mse\_chat\_room\_registry} actor can take part in multiple instances of the \texttt{ChatServer} protocol to connect multiple different clients to chat rooms. 
Instead, in \mse{}, actors are associated with roles using a configuration file. 

\paragraph{Process Structure} When working in the setting of a functional, actor-based language with immutable variables, it is convenient to have separate processes for monitoring and user logic (Figure~\ref{fig:mse-session-actor}).

The solution used in \mse{} is to provide a \emph{session key} or \texttt{ConvKey} to the user. A session key is a 3-tuple $(M, R, S)$ where $M$ is the process ID of the monitor process, $R$ is the name of the role that the participant is playing in the current message handler, and $S$ is the process ID of the coordinating process for the current session. To the user, the session key is an opaque, abstract value. Passing the session key as a parameter to the send operation allows the correct monitor to be identified in order to check the outgoing message and update the monitor state. The monitor state will be updated in the monitoring process, and consequently there is no requirement for linearity tracking. 

\subsection{Session Initiation}
\label{sec:initiation}

\begin{figure}[htpb!]
\centering
\includegraphics[width=\linewidth]{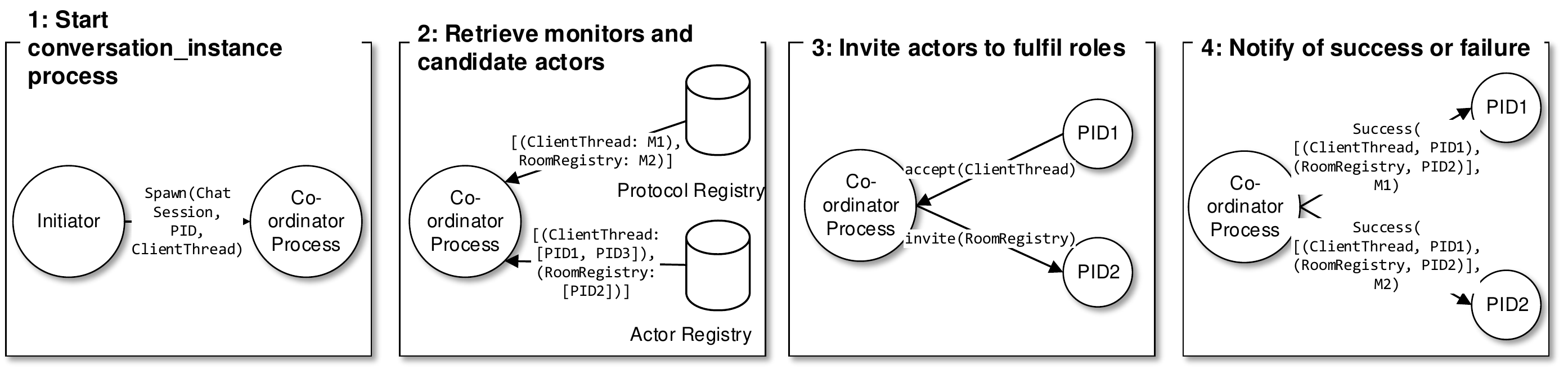}
\caption{Actor Invitation Workflow}
\label{fig:design-impl:invitation}
\end{figure}

A strength of the multiparty session actor framework is that messages can be sent to a role name instead of requiring a concrete actor address, but this introduces the issue of associating roles with endpoints at the beginning of the protocol. 
The work of~\citet{neykova:actors} makes essential use of the Advanced Message Queueing Protocol (AMQP), which provides abstractions known as \emph{exchanges} to distribute messages to other AMQP entities. Session initiation requires four exchanges, and all session communication is routed through a single exchange per session instance.

%
\noindent
As Erlang is an actor-based language, we do not use AMQP, resulting in a simpler invitation workflow. We require two centralised registries: a \texttt{protocol\_registry}, which contains precomputed monitor FSMs for each role in a protocol, and an \texttt{actor\_registry}, which associates active session actors with the roles they may fulfil. More concretely, the \texttt{actor\_registry} is a map $\textit{Protocol Name} \mapsto (\textit{Role} \mapsto \textit{Actor PID})$.
%
The procedure for initiating a session (shown in Figure~\ref{fig:design-impl:invitation}) is as follows:
\begin{enumerate}
\itemsep-0.25em
\item A session actor---the \emph{session initiator}---requests that a session is initiated, specifying a protocol name, and the role it wishes to take in the protocol.
\item A \texttt{conversation\_instance} process is spawned to co-ordinate session actions.
\item The \texttt{conversation\_instance} process contacts the \texttt{protocol\_registry} process to retrieve the list of roles and monitors used in the process, and contacts the \texttt{actor\_registry} process to retrieve the list of actor process IDs which may fulfil the roles in the session.
\item The \texttt{conversation\_instance} process invites eligible actors to fulfil each role.
Once all roles have been fulfilled, each actor is notified, invoking the \texttt{ssactor\_conversation\_established} callback.
If it is not possible to fulfil a role, for example because all active session actors have declined the invitation to fulfil the role, then the invitation process is aborted. All actors already invited to fulfil roles in the protocol are notified, resulting in the invocation of the \texttt{ssactor\_conversation\_error} callback.
\end{enumerate}

\subsection{Monitoring}

Messages are checked against monitors based on communicating finite-state machines~\cite{brand:cfsms}. Firstly, global types are projected into local types, and the local types are used to construct monitors based on communicating finite state machines using an algorithm based on that of~\citet{denielou:automata}. 

\begin{wrapfigure}[13]{l}{0.35\textwidth}
\includegraphics[width=\linewidth]{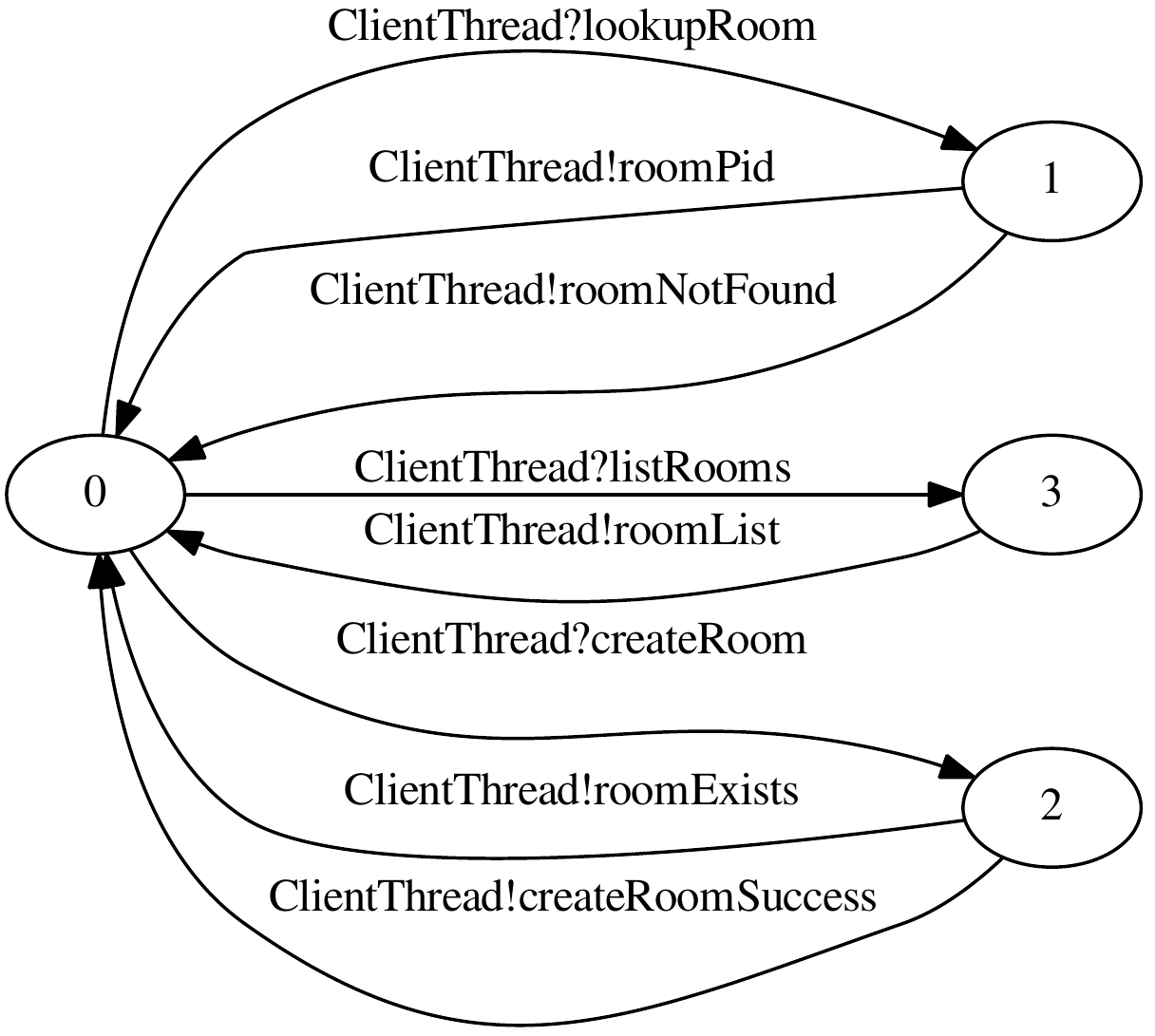}
\caption{Monitor for ClientThread role of ChatServer protocol}
\label{fig:clientthread-monitor}
\end{wrapfigure}

Transitions between states are predicated on send and receive operations. The monitor generated for the \texttt{ChatServer} protocol from Section~\ref{sec:overview:chat-server} projected at the \texttt{RoomRegistry} role is shown in Figure~\ref{fig:clientthread-monitor}.

The monitor generation algorithm also makes use of the nested FSM optimisation described by~\citet{hu:interruptible} to ensure that generated monitors are polynomial in the size of the global type in the presence of parallel composition. 
Instead of generating states for all possible interleavings, the algorithm generates a separate monitor for each block of interactions composed in parallel, and the outer monitor can only progress when all nested monitors are in a terminal state.

A notable difference to standard CFSMs is that, following the design of Scribble, we allow transitions to be predicated on sending a message to a \emph{set} of recipients, in a multicast fashion. 
The failure detection mechanisms in Section~\ref{sec:failure-handling:failure-detection} ensure all recipients accept the message. 

Monitor generation takes place when the actor system is started, and the generated monitors are stored in the \texttt{protocol\_registry} actor. 
The second aspect of monitoring is the monitoring runtime: once a monitor has been generated, it may be used to check incoming and outgoing messages against the local specification for a type. 
The monitor process for an actor contains a hashtable mapping session ID / role pairs to monitors.
Checking a message against a monitor involves checking whether any transitions can be made from the current monitor state. If not, then an exception is raised.

\subsection{Sending Messages}
A user may send a monitored session message by calling the \texttt{conversation:send} function. In order to send a message, four pieces of information are required: a session key \texttt{ConvKey}; a list of recipients; a message name; and a list of values. Recall that the monitoring process is \emph{external} to the logic process, and the session key contains the role name and the session ID, which uniquely identify a monitor.

When \texttt{conversation:send} is called, a message is sent to the actor monitor process, which retrieves the appropriate monitor using the role name and session ID. Should the message be accepted by the monitor, the role will be resolved to the PID of the monitor of the receiving process by the routing table. At this point, a synchronous call will be made to the remote monitor to ascertain whether the message can be accepted: if so, both monitors are advanced, and the actor will proceed. 
The monitoring process is synchronous, in order to allow errors to be reported to the user. 

\emph{If a monitor rejects a message, an exception is thrown. Such behaviour is consistent with the Erlang design ideology of letting a process fail if its behaviour deviates from a specification. As a result, \mse{} extends the let-it-fail ideology to communication patterns.}

\subsection{The \texttt{ssa\_gen\_server} Behaviour}
\label{sec:design-impl:ssa-gen-server}
\begin{figure}[t]
\centering
\includegraphics[scale=0.725]{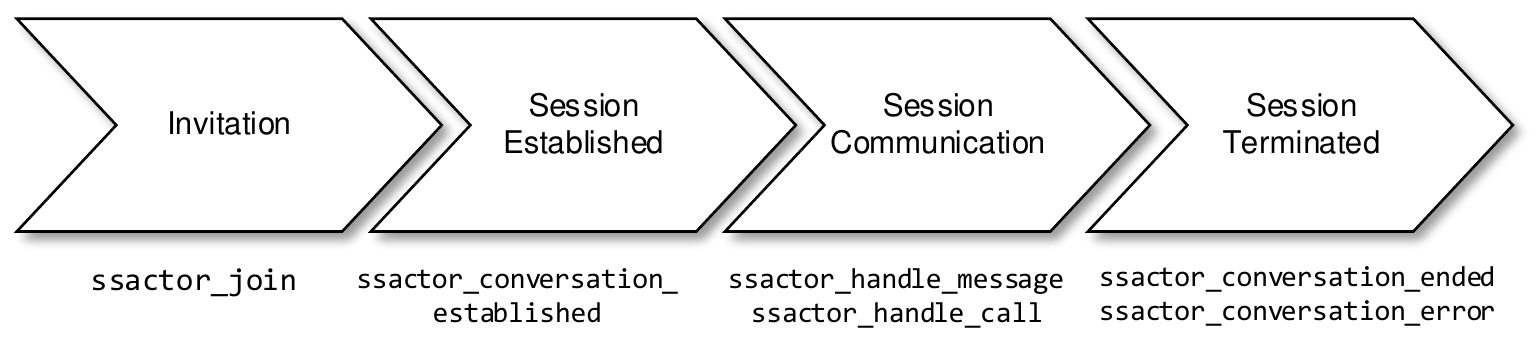}
\caption{Session Lifecycle}
\label{fig:session-lifecycle}
\end{figure}
In keeping with the Erlang/OTP method of designing applications, \mse{} provides a behaviour, \texttt{ssa\_gen\_server}, which contains callbacks that should be implemented by Erlang session actors.
Figure~\ref{fig:session-lifecycle} describes the session lifecycle, and the expected callbacks.

During session initiation, actors eligible to fulfil a role will be invited to participate in the session, triggering the \texttt{ssactor\_join} callback. The expected return value consists of a pair of either the atom \texttt{accept} or \texttt{decline},  and an updated actor state. 
Once all roles in the protocol have been fulfilled, the \texttt{ssactor\_conversation\_established} callback is invoked. At this point, the actor can begin to communicate using session messages. When a message is received, \texttt{ssactor\_handle\_message} is invoked.

Once all communication has finished an actor may end the session, which invokes \texttt{ssactor\_conversation\_\ ended} in all participant sessions. Alternatively, \texttt{ssactor\_conversation\_error} is invoked if an error occurs and the session cannot continue (for example, an actor that is required in the session terminates).

\section{Failure Detection and Handling}
\label{sec:failure}
A common assumption for implementations of either session-typed languages, or monitoring frameworks for applications using session types, is that processes persist throughout the course of the session. 

Unfortunately, this assumption does not hold true in Erlang applications. An important design pattern in Erlang applications is to arrange processes in \emph{supervision hierarchies}, allowing processes to fail and be restarted by their supervisors when they encounter an unrecoverable fault. Consequently, it is not possible to assume that a process is running throughout the entirety of the session. In this section, we detail how failures within a session can be detected, the circumstances in which a session can continue in spite of the termination of a participant, and a modular method based on \emph{subsessions} to facilitate error handling.

To do so, we rely on two tools provided by Erlang: firstly, Erlang provides the possibility to emulate \emph{synchronous calls}, with a call returning an error or timing out if a remote actor has terminated or is unreachable; and secondly, Erlang provides the possibility to register for reliable \emph{notifications} should a process terminate, known as \emph{monitoring} processes. The latter method, not to be confused with monitoring the communication between actors, is reliable both in the case of a single node, and in the case that a remote node becomes unreachable.

\subsection{Failure Detection}
\label{sec:failure-handling:failure-detection}
Should a process in the session fail, the failure should be detected, as it may be the case that the process which has failed is playing a role which is involved in the remainder of the session.

To this end, we describe two methods of failure detection: push-based, which involves using the Erlang \texttt{monitor} functionality to detect when a participant is no longer available, and pull-based, which uses reliable sends and a two-phase commit protocol.
\paragraph{Push-Based}
\label{sec:failure-handling:failure-detection:push}

\begin{figure}
\centering
\includegraphics[width=\linewidth]{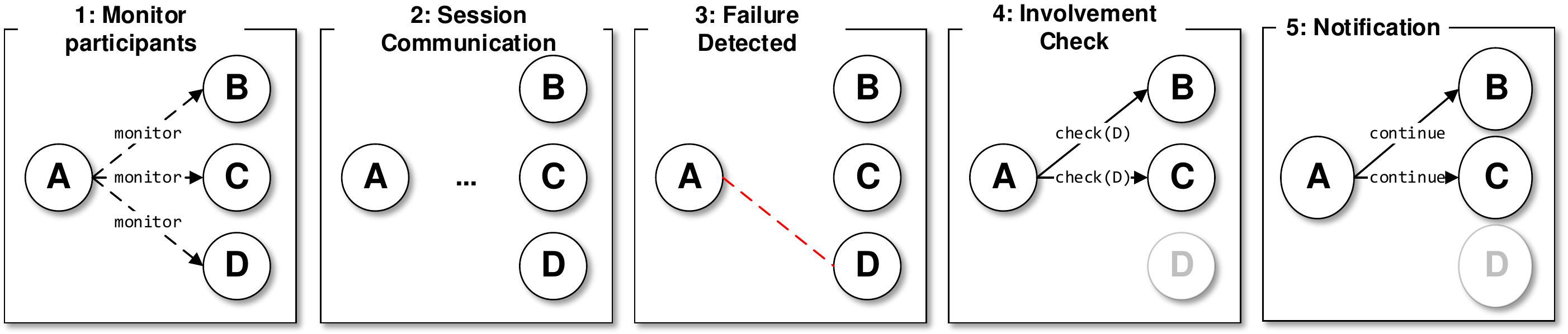}
\caption{Push-Based Failure Detection}
\label{fig:failure-handling:push-based-diag}
\end{figure}

Push-based failure detection uses Erlang's reliable termination detection functionality to notify other participants in the session that an unrecoverable failure has occurred. 
Figure~\ref{fig:failure-handling:push-based-diag} shows the main stages of the push-based failure detection system: firstly, the co-ordinator process for a session monitors each participant in the session to be notified if any of the processes terminate.
Should a failure in any of the processes be detected, then the co-ordinator process sends a request to each other participant of the session to determine whether, from the point of view of each participant, the terminated actor is involved in the remainder of the session. More specifically, a role $r$ is \emph{involved} in a session if there exists a transition reachable from the current monitor state where $r$ is the sender or receiver in a communication.

After the involvement check is complete, then participants are notified of the result. Should all actors respond that the terminated participant is not involved in the remainder of the session, then the session may continue as before; if the terminated participant is involved in the remainder of the session or any of the other actors in the session are unreachable, however, then it is not possible to continue, and the remaining participants are notified of the session's termination.

\begin{figure}[htpb]
\centering
\begin{subfigure}[c]{0.7\linewidth}
\includegraphics[width=\linewidth]{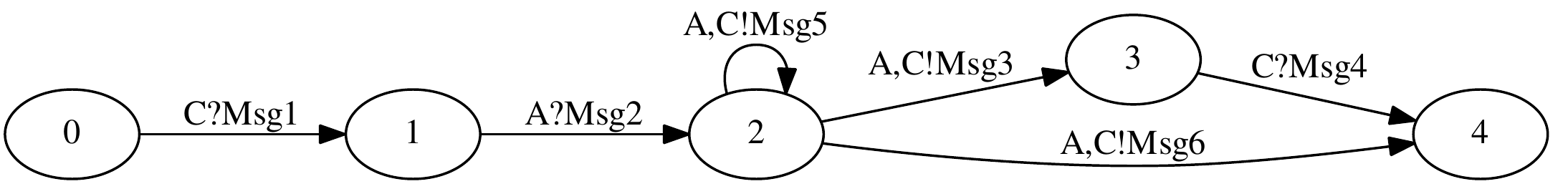}
\caption{Example Monitor}
\label{fig:failure-handling:two-buyer-buyer2}
\end{subfigure}
\begin{subfigure}[c]{0.25\linewidth}
\footnotesize
\centering
\begin{tabular}{|l|l|} \hline
State & Reachable Roles \\ \hline
0 & [A, C] \\ \hline
1 & [A, C] \\ \hline
2 & [A, C] \\ \hline
3 & [C] \\ \hline
4 & [] \\ \hline
\end{tabular}
\caption{Reachability table}
\label{tbl:failure-handling:tbp-roles}
\end{subfigure}
\caption{Monitor Reachability Analysis}
\label{fig:monitor-reachability}
\end{figure}
\noindent
Push-based failure detection includes a check to see whether a given role is involved in the remainder of the session. 
Figure~\ref{fig:failure-handling:two-buyer-buyer2} shows a monitor and a table detailing the roles reachable at each state\footnote{Some readers may recognise this monitor as that of buyer 2 in the two-buyer protocol~\cite{honda:multiparty}.}; the reachability algorithm only needs to be run once, upon monitor generation, and also records the IDs of any nested FSMs used in order to detect roles inside \texttt{par} blocks. Checking whether a role is involved at the current point in the session is achieved by a lookup of the current state ID.

\paragraph{Pull-Based}

\begin{wrapfigure}[21]{l}{4.5cm}
\footnotesize
\centering
\begin{subfigure}[t]{\linewidth}
\centering
\includegraphics[width=\linewidth]{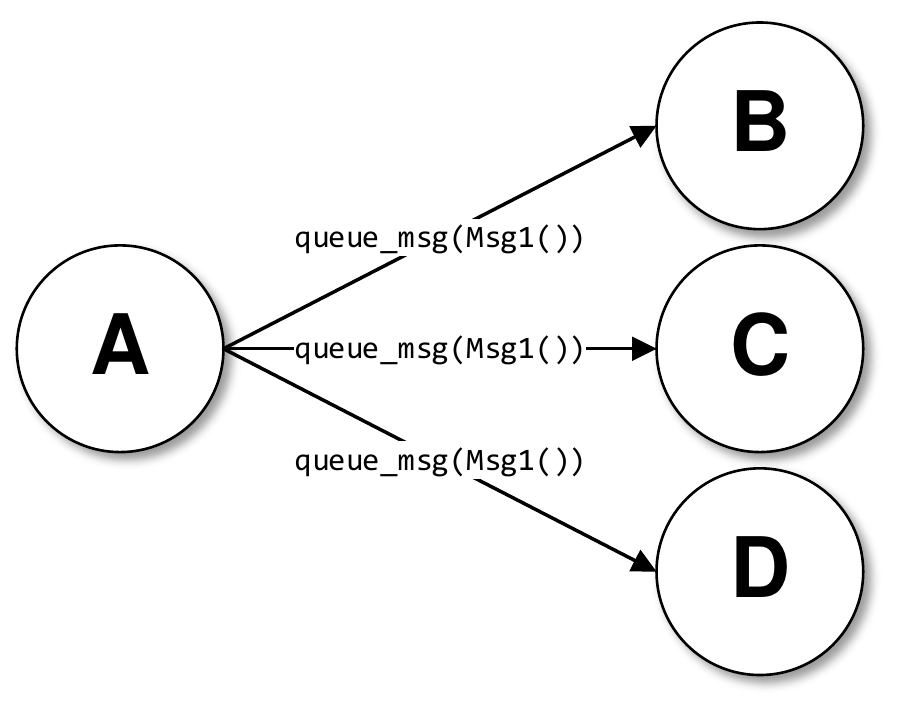}
\caption{Queue Messages: Successful}
\label{fig:failure-handling:pull-queue}
\end{subfigure}

\begin{subfigure}[t]{\linewidth}
\centering
\includegraphics[width=\linewidth]{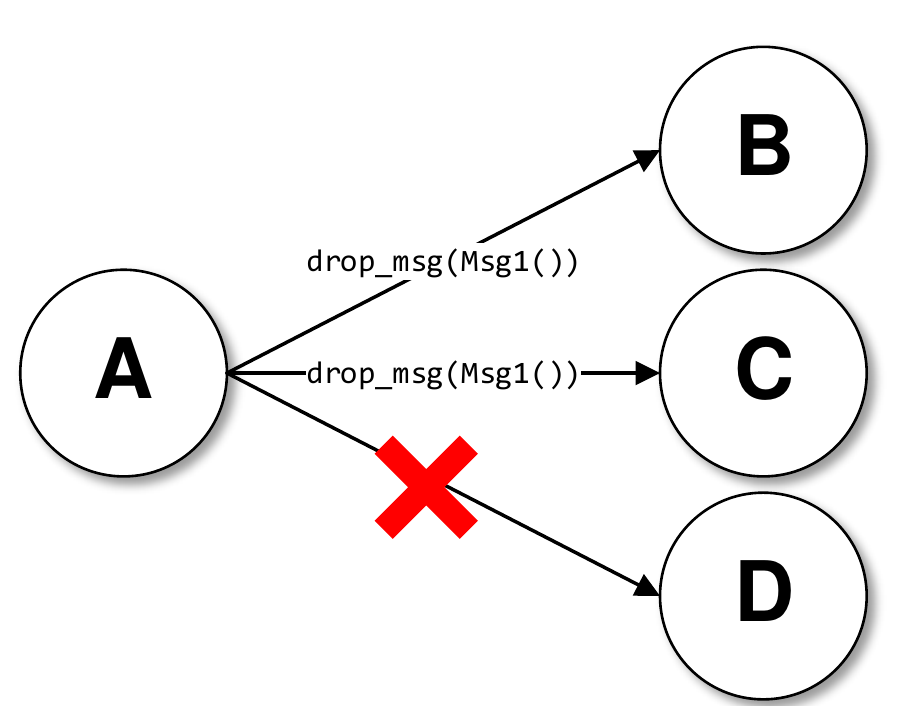}
\caption{Queue Messages: Failure}
\label{fig:failure-handling:pull-queue-fail}
\end{subfigure}
\caption{Pull-based failure detection}
\label{fig:failure-handling:pull-queue-fig}
\end{wrapfigure}

As Scribble protocols allow messages to be sent to multiple participants, it is desirable to ensure that messages are only delivered if all processes receiving the message are active. 
To do so, pull-based failure detection uses a two-phase commit to ensure that all recipient processes are available, and that all recipient monitors accept the message.

The first stage of pull-based failure detection is to send a synchronous message, \texttt{queue\_msg}, to each recipient monitor process (Figure~\ref{fig:failure-handling:pull-queue}). Either the call will succeed, returning \texttt{ok}, indicating that the call was successful and the message was accepted by the remote monitor; the call will succeed, but returning \texttt{error}, indicating that the remote process was available but the remote monitor rejected the incoming message; or the call will fail.
When a message is queued, it is assigned a unique identifier. Should a message be accepted, it is stored in a table; should all messages be delivered successfully, a second, asynchronous message will be sent to \emph{commit} the message.


If the a queue message fails for any participant however (Figure~\ref{fig:failure-handling:pull-queue-fail}), then the message cannot be delivered successfully. If the failure is due to a message rejection, then it is possible for the session to continue: a \texttt{drop} message is sent to all participants, the messages are discarded from the queue, and the failure is synchronously reported to the sender. The session cannot continue if a process is unreachable.


\paragraph{Discussion}
Push-based approaches allow failures to be detected as soon as they occur, and allow the sessions to continue should the failed role not be involved in the remainder of the session.
Pull-based approaches only report failures when a failed role is needed, but do not require co-ordination amongst processes to detect whether it is safe to continue. 
On the other hand, however, pull-based detection approaches fall short when an actor terminates while processing a message; consider 
a protocol \lstinline+X() from A to B, C; Y() from B to A, C;+.

Suppose \texttt{X} is delivered successfully, but B terminates prior to sending \texttt{Y} to \texttt{A}: in this case, there would be no way of detecting the failure. Such a situation can be detected using push-based detection. Pull-based detection also falls short if a process terminates between queueing and committing a message.
\noindent
Push-based failure detection falls short should a message handler involving the failed role be executed while the safety check is in progress. Suppose \texttt{A} terminates while \texttt{B} processes message \texttt{X}. Without pull-based detection, there no guarantee that the failure will be detected before \texttt{Y} is sent to A and C, with only C receiving the message.
Consequently, it is useful to use both methods of failure detection together to ensure that failures are eventually detected (using push-based detection) and that they are detected should the process fail before the safety check is complete (using pull-based detection).

\subsection{Subsessions for Exception Handling}
In unpublished work\footnote{\url{http://www.doc.ic.ac.uk/~rn710/sactor/main.pdf}}, Neykova and Yoshida apply \emph{subsessions} to dynamically introduce actors into roles, in particular demonstrating the technique using a Fibonacci benchmark. 
The session actor framework requires all roles in a protocol to be fufilled upon session initiation, but due to the common Erlang practice of storing process IDs in registries and passing them in messages, it is often the case that it is not known which actor should fulfil which role until later in the protocol. As an example, consider again the chat server: a user sends a room name to the room registry which, if the room exists, responds with the room's process ID. It is only at this point that we know which actor should fulfil the \texttt{ChatRoom} role! 

Subsessions are a modular abstraction which allow such a pattern to be encapsulated. Interestingly, at the end of their paper on nested protocols, \citet{demangeon:subsessions} state:

\begin{displayquote}
\small
\textit{
Yet exceptions are absolutely necessary when specifying real-world protocols. We believe that nested protocols give a simple way to handle exceptions, by making explicit blocks of computation.}
\end{displayquote}
\noindent
This is especially poignant given a setting where actors can terminate in the middle of a session: by splitting protocols into subprotocols, it is possible to repeat parts of a session with possibly-different participants should a participant in a subsession terminate. 

Demangeon and Honda also describe a method by which subsessions may return results, which we can adapt to implement a simple failure handling mechanism. We allow participants in a subsession to state that the subsession has failed through the \verb+conversation:subsession_failed+ function, which takes an argument stating the failure. We can then introduce the \texttt{initiates} construct:

\begin{lstlisting}
InitiatorRole initiates ProtocolName(Roles) { SuccessBlock } handle(FailureName) { FailureBlock }
\end{lstlisting}

This construct states that role \texttt{InitiatorRole} initiates an instance of the \texttt{ProtocolName} protocol as a subsession. If the subsession completes successfully (that is, if the session finishes and an actor calls \verb+conversation:subsession_complete+), then the protocol proceeds as \texttt{SuccessBlock}. If the process calls \verb+conversation:subsession_failed+ with the argument \texttt{FailureName} however, then the protocol proceeds as \texttt{FailureBlock}. A process can have an unlimited number of \texttt{handle} clauses. If no failures are expected, then \texttt{initiates} can be used without an explicit \texttt{SuccessBlock}.

The \texttt{initiates} construct indicates that the session initiator initiates the session, and makes a choice based on the result.
To every role other than the initiator, the construct is simply projected as a choice directed by the initiator role: safety follows from enforcing the same restrictions on \texttt{SuccessBlock} and each \texttt{FailureBlock} as in Scribble choice blocks. We provide a distinguished reason for failure, \texttt{ParticipantOffline}, which is returned should a session be aborted due to failure detection.

\begin{wrapfigure}[9]{l}{6cm}
\vspace{-1em}
\begin{lstlisting}[frame=none]
ClientThread initiates ChatSession(ClientThread, 
    new ChatRoom) {
  clientLeftRoom() from ClientThread to Logger;
  continue ClientChoiceLoop;
} handle(Kicked) {
  clientKicked() from ClientThread to Logger;
  continue ClientChoiceLoop; 
} handle(ParticipantOffline) {
  roomTerminated() from ClientThread to Logger;
  continue ClientChoiceLoop; 
}
\end{lstlisting}
\vspace{-1em}
\end{wrapfigure}

As an example, let us introduce a \texttt{Logger} process to the \texttt{ChatServer} protocol, and state that a client thread should send a message to the logger with the reason for leaving. 
We introduce two \texttt{handle} clauses: if the subsession ends with \texttt{Kicked}, then a moderator has expelled the client from the room, whereas if the subsession ends with \texttt{ParticipantOffline} is called, then the actor playing the \texttt{ChatRoom} role has terminated. In both cases, the appropriate messages are sent to the logger, and the client is free to join another room.
\vspace{-2em}

\section{Evaluation}

\subsection{DNS Server Case Study}
\begin{figure}
\begin{subfigure}{0.5\linewidth}
\includegraphics[width=0.9\linewidth]{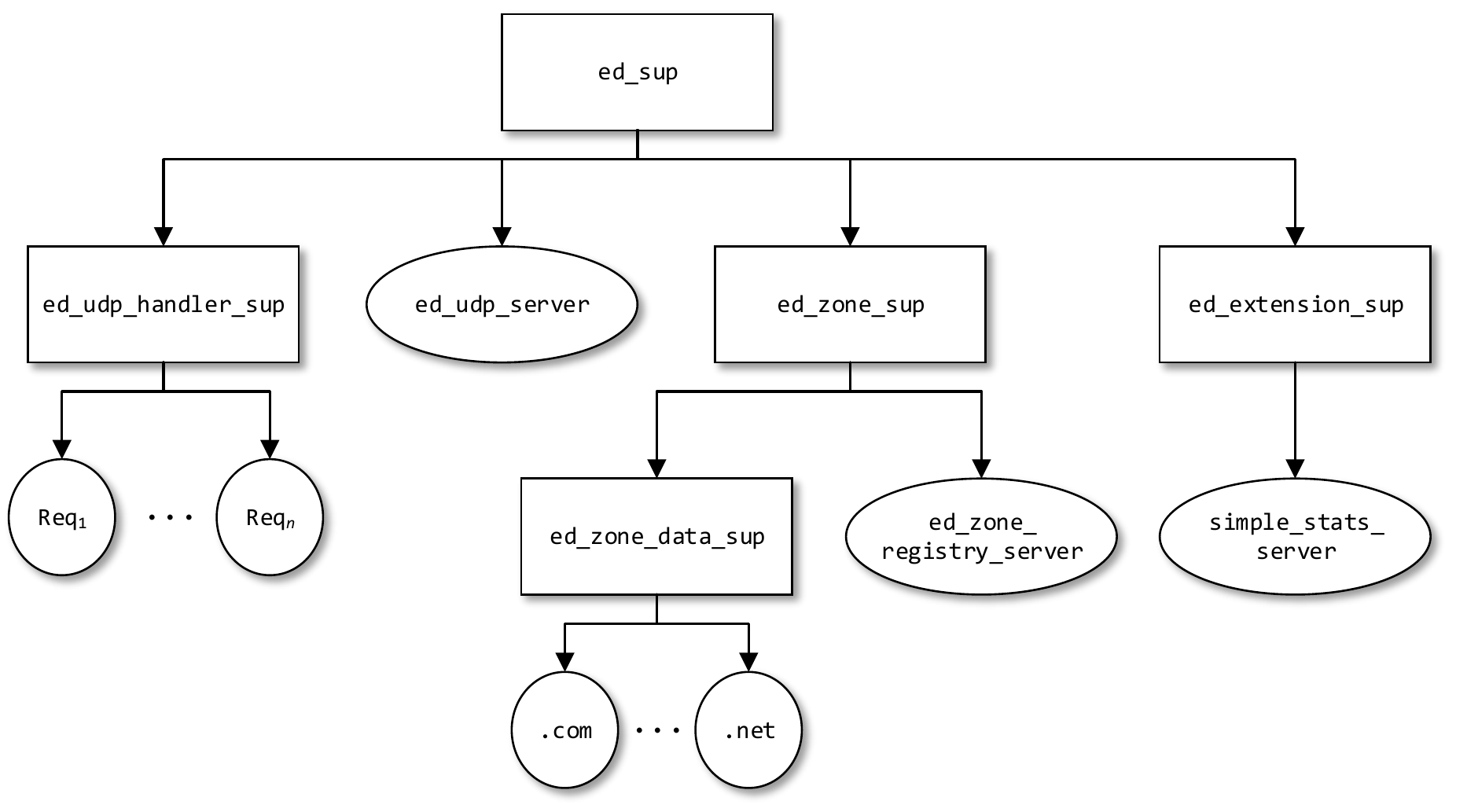}
\caption{Supervision hierarchy for \texttt{erlang-dns}}
\label{fig:eval:erlang-dns-hierarchy}
\end{subfigure}
~
\begin{subfigure}{0.5\linewidth}
\includegraphics[width=0.9\linewidth]{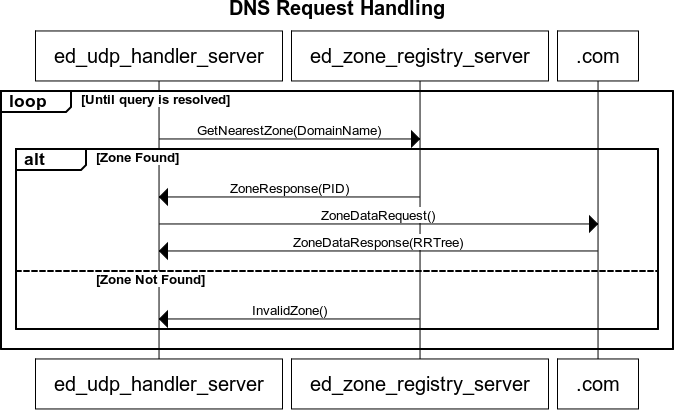}
\caption{Messages sent to fulfil a DNS lookup request}
\label{fig:eval:erlang-dns-messages}
\end{subfigure}
\caption{DNS Server Case Study}
\end{figure}
\noindent
The \texttt{erlang-dns} project\footnote{\url{https://www.github.com/hcvst/erlang-dns}} is an Erlang/OTP server for the Domain Name System (DNS). Figure~\ref{fig:eval:erlang-dns-hierarchy} shows the supervision hierarchy of the server: 
of interest to protocol are zone data servers (shown as `.com' and `.net'), which map domain names to IP addresses; \texttt{ed\_zone\_data\_server}, which maps domain names to zone data servers; and UDP handler servers (shown in the diagram as $\texttt{Req}_i$), which handle requests.

Upon system initiation, the \texttt{ed\_udp\_server} process opens a UDP acceptor socket and listens for incoming requests. When a query is received, an \texttt{ed\_udp\_handler\_server} process is spawned to handle the request, and it is at this point that the session is started. 
Figure~\ref{fig:eval:erlang-dns-messages} shows the messages sent when fulfilling a DNS request: the UDP handler server contacts the zone registry to ascertain whether or not the zone exists. If not, the server returns \texttt{InvalidZone}, at which point a DNS response packet to this effect is generated and sent back to the client. If the zone is found, the zone registry returns the PID of the zone server, which is queried for information about the zone. At this point, if the IP address can be resolved from the request, then it is returned to the user, but it may also be necessary to perform a recursive lookup.

\paragraph{Protocol}
Figure~\ref{fig:edns-protocols} shows the Scribble protocols for \texttt{erlang-dns}. There are three roles: \texttt{UDPHandlerServer}, fulfilled by the \texttt{ed\_udp\_handler\_server} instance initiating the session; \texttt{DNSZoneRegServer}, fulfilled by \texttt{ed\_zone\_registry\_server}; and \texttt{DNSZoneDataServer}, fulfilled by a zone data server able to handle the request. It is not possible to fulfil the \texttt{DNSZoneDataServer} role upon session initiation as the actor to invite depends on the result of the request to the zone registry; consequently, we have a main protocol \texttt{HandleDNSRequest}, and a subprotocol \texttt{GetZoneData} which is initiated should \texttt{ed\_zone\_registry\_server} respond with a PID.

\begin{figure}[t]
\begin{lstlisting}[multicols=2, frame=none, numbers=left,xleftmargin=3em] 
global protocol HandleDNSRequest(role UDPHandlerServer, role DNSZoneRegServer) {
  rec QueryResolution {
    FindNearestZone(DomainName) 
      from UDPHandlerServer to DNSZoneRegServer;
    choice at DNSZoneRegServer {
       ZoneResponse(ZonePID) from DNSZoneRegServer to UDPHandlerServer;
       UDPHandlerServer initiates GetZoneData(
         UDPHandlerServer, new DNSZoneDataServer);
       continue QueryResolution;
     } or { InvalidZone() from DNSZoneRegServer to UDPHandlerServer; } 
  } 
} 

global protocol GetZoneData(role UDPHandlerServer, role DNSZoneDataServer) {
  ZoneDataRequest() from UDPHandlerServer to DNSZoneDataServer;
  ZoneDataResponse(RRTree) from DNSZoneDataServer to UDPHandlerServer;
}
\end{lstlisting}
\vspace{-.75em}
\begin{lstlisting}[language=erlang, frame=none,basicstyle=\scriptsize\ttfamily]
config() ->
  [{ed_zone_data_server, [{"GetZoneData", ["DNSZoneDataServer"]}]},
   {ed_zone_registry_server, [{"HandleDNSRequest", ["DNSZoneRegServer"]}]},
   {ed_udp_handler_server, [{"HandleDNSRequest", ["UDPHandlerServer"]}, {"GetZoneData", ["UDPHandlerServer"]}]}].
\end{lstlisting}
\vspace{-1em}
\caption{Scribble Protocol and Configuration File for \texttt{erlang-dns}}
\label{fig:edns-protocols}
\end{figure}
\paragraph{Implementation}
To adapt \texttt{erlang-dns} to use \mse{}, we firstly define a configuration file to associate each actor with the role that it plays in each protocol. 
Next, we adapt each of the actors which were formerly instances of \texttt{gen\_server} to be instances of \texttt{ssa\_gen\_server}. 
Vitally, \emph{no changes are made to the supervision structure}, as it is \emph{orthogonal} to the monitoring of messages. 

\begin{wrapfigure}[17]{l}{7.7cm}
\centering
\begin{subfigure}{\linewidth}
\centering
\includegraphics[width=\linewidth]{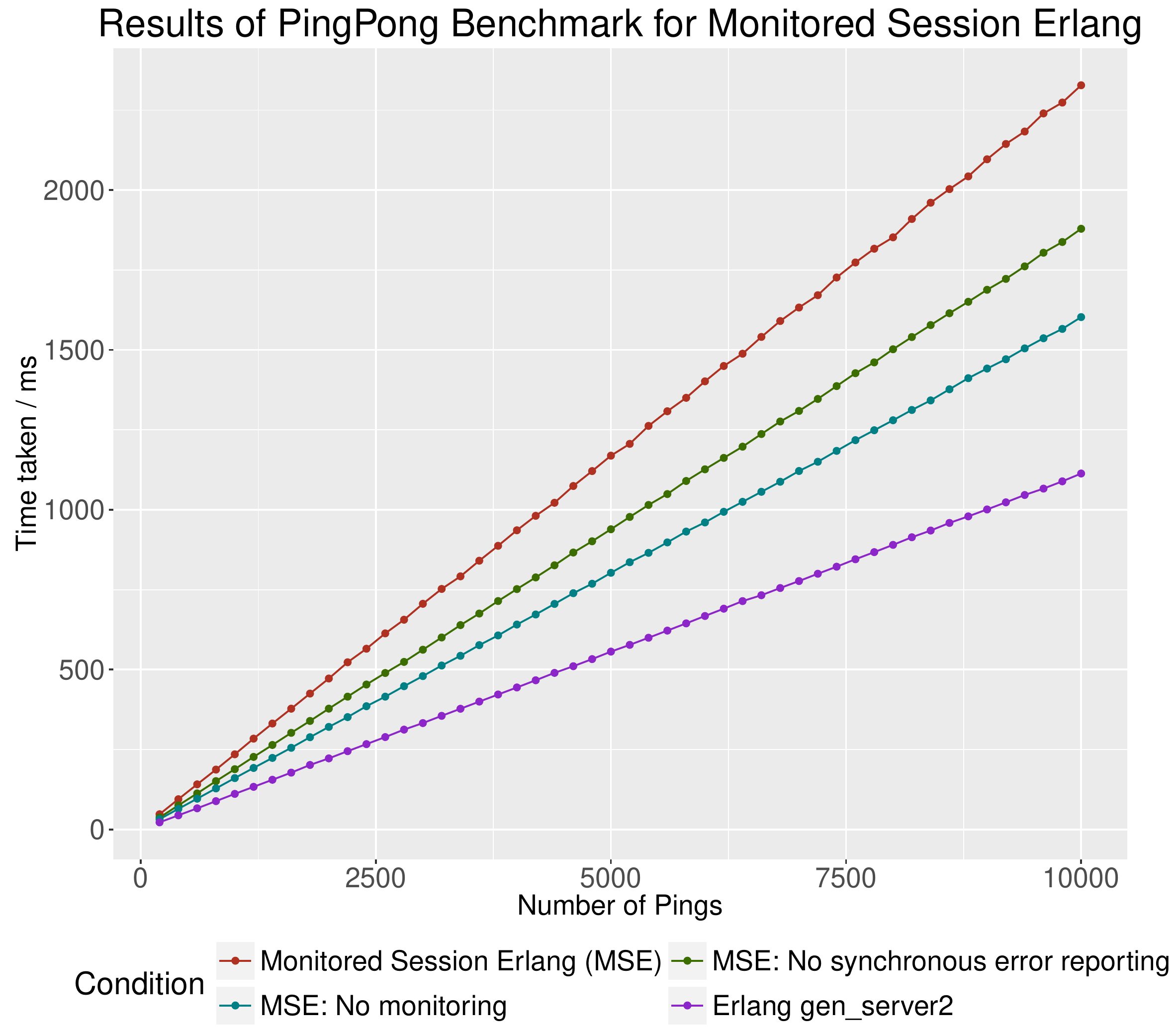}
\caption{PingPong experimental results}
\label{fig:eval:pingpong-graph}
\end{subfigure}
\caption{Experimental evaluation of overheads}
\end{wrapfigure}

There are two main ways that the session implementation diverges from the original implementation. Firstly, once the zone data server PID has been returned, we start a new subsession to invite the zone data server and retrieve the zone data. Secondly, we emulate a synchronous call with two asynchronous messages.
%
As DNS is most commonly implemented over UDP, there is no guarantee that a response will ever be received. Should a failure occur within the DNS server, there is little point in trying to fulfil the remainder of a request. Instead, it is better to allow the supervisor to restart the component and let the request time out.

\subsection{Overheads} \label{sec:eval-overheads}

We measure the overheads of the framework using the \texttt{PingPong} benchmark: an actor \emph{A} sends a message \texttt{ping} to an actor \emph{B}, which responds with a message \texttt{pong}. 
We measure several scenarios: ``Erlang \texttt{gen\_server2}'' refers to an implementation not using \mse{}, where actors communicate using the \texttt{gen\_server:cast} function. The remaining three scenarios use \mse{}: ``MSE'' refers to an implementation using the full system; ``MSE: No synchronous error reporting'' refers to an implementation without synchronous reporting of monitoring errors, and ``MSE: No monitoring'' refers to an implementation without monitoring.

The different scenarios demonstrate different aspects of the system: 
in contrast to the original work on session actors, 
sending a message involves synchronous calls to both the source and destination monitors to immediately report failures. 

Should either of the checks fail, 
an exception is raised, allowing the computation to be aborted as soon as a message is rejected by a monitor.
Employing an asynchronous approach results in two fewer messages, but errors have to be reported as separate messages, meaning that the remainder of the handler must run before an error report can be processed. 
The final variation uses \mse{} with monitoring disabled: the overheads incurred in this scenario are as a result of the external monitor process and the resolution of role names to actors.

%

Figure~\ref{fig:eval:pingpong-graph} shows the experimental results of the four basic experimental scenarios\footnote{Experimental conditions: Two cluster nodes with 4 16-core AMD Opteron 6376 processors at 2300MHz; 264GB RAM; RTT 0.101ms using \texttt{ping}. Scientific Linux 7, Erlang 7.0. Value plotted: arithmetic mean over 100 repetitions, measured after session establishment.}. The \texttt{gen\_server2} implementation is fastest at 0.111ms per iteration, whereas the full \mse{} system has a mean time per iteration of 0.23ms, giving a final overhead per iteration of 0.12ms (or 0.06ms per messsage). The overheads can be explained by the additional messages sent between the monitors in order to detect and report monitoring errors.

\section{Related Work}
Session types were originally introduced by~\citet{honda:dyadic} and later expanded upon by~\citet{honda:primitives} to model interactions between two communicating parties. 
%
\citet{honda:multiparty} propose multiparty session types to model interactions with more than two participants. A \emph{global type} is projected to a \emph{local type} for each participant. Conformance to local types can be checked statically, or monitored at runtime.

\citet{denielou:automata} discuss the connections between multiparty session types and communicating finite state machines~\cite{brand:cfsms}: we use a variant of this algorithm in the monitor generation phase. \citet{chen:asynchronous} and~\citet{bocchi:monitoring} formalise the theory of runtime monitoring, with monitors and routing tables as first class entities, and reductions of monitored processes predicated on labels emitted by a labelled transition system on local types.  
The SPY framework \cite{neykova:spy} can monitor communication against multiparty session types in Python, and runtime monitoring can also be used to enforce timing constraints~\cite{neykova:timed}. 

\citet{capecchi:escape} describe a process calculus with multiparty session types containing \verb+try..catch+ and \verb+throw+ constructs, where a set of roles move to a compensation process should an exception be thrown. 
\citet{chen:robust-tt} describe a formal system of \emph{protocol types} based on multiparty session types, where each interaction is annotated with the exceptions that may occur as a result of the interaction, and continuations which are invoked upon a failure occurring. The system is realised by a \emph{transformation} stage which combines projection with an analysis of participants which need to be notified should an exception occur. The formalism is elegant and a particular strength is its decentralised nature. In our setting, the requirement to satisfy the protocol well-formedness conditions for choice blocks means that such notifications must be encoded explicitly in the protocol. In contrast, our approach of using subsessions to structure protocols allows parts of the protocol to be retried with different participants if one terminates or goes offline.

\citet{mostrous:session-erlang} consider a core session calculus based on Erlang, using Erlang's ability to generate fresh references to design a static type discipline relying on \emph{correlation sets}~\cite{viroli:correlation-sets} to associate messages with sessions. The type system has not yet been implemented. \citet{crafa:actors} defines a behaviourally-typed actor caclulus, AC, based on Scala's actor primitives, guaranteeing that each input is eventually matched with exactly one output in the system. 
%
We expand upon the work of~\citet{neykova:actors} in monitoring actors according to multiparty session types by investigating how the session actor framework can be applied in the setting of an actor-based functional language without relying on AMQP, and propose methods of handling the case where an actor terminates during a session.

 
\vspace{-0.5em}
\section{Conclusion}
Communication is central to software written in actor-based languages. We have described the design and implementation of \mse{}, a framework that allows communication in Erlang applications to be monitored against multiparty session types. Our tool demonstrates the applicability of the conceptual framework of multiparty session actors to actor-based functional languages, motivates subsessions as a fundamental abstraction, and introduces features such as synchronous error reporting and allowing actors to take part in multiple instances of a protocol.

Future work will centre around a formal semantics, and an investigation into how ideas from the dynamic view of session-typed actors can be used in a session-typed concurrent $\lambda$-calculus.
\vspace{0.3em}

%
%

\small
\paragraph{Acknowledgements}
Thanks to Sam Lindley, Garrett Morris, and Philip Wadler for useful discussions, and to the anonymous reviewers for detailed and insightful comments. This work was supported by EPSRC grant EP/L01503X/1 (University of Edinburgh CDT in Pervasive Parallelism).
\vspace{-0.5em}
\bibliographystyle{plainnat}
\setlength{\bibsep}{0pt plus 0.3ex}
\footnotesize
\bibliography{refs}
\end{document}